\documentclass{PoS}

\title{Preliminary results of $\Delta I=1/2$ and $3/2$, $K$ to $\pi\pi$ Decay Amplitudes from Lattice QCD}

\ShortTitle{$K$ to $\pi\pi$ Decay Amplitudes}

\author{\speaker{Qi Liu}\\
        Department of Physics, Columbia University, New York, NY 10025, USA \\
        E-mail: \email{ql2142@columbia.edu}}

\author{RBC and UKQCD collaborations}
\abstract{
We report a direct lattice calculation of the $K$ to $\pi\pi$ decay matrix elements for both $\Delta I=1/2$ and $3/2$ channels on 2+1 flavor, domain wall fermion, $16^3\times32$ lattices with zero $\pi\pi$ relative momentum and $m_\pi=420$ MeV. All $K^0$ to $\pi\pi$ contractions are carefully listed and calculated.  The decay into the isospin zero $\pi\pi$ final state, which receives contributions from the disconnected graphs, is very difficult to calculate, but a clear signal in the similar disconnected $\pi\pi$ correlator can be seen. 
 Preliminary results, some with large errors, will be presented for the various contributions to the renormalized weak matrix elements $A_0$ and $A_2$. We obtain Re$(A_0)$ with $25\%$ error in the case of zero momentum on shell decay, and find a factor of 6 enhancement for the $\Delta I=1/2$ rule in the $420$ MeV pion system.
}

\FullConference{The XXVIII International Symposium on Lattice Field Theory, Lattice2010\\
		June 14-19, 2010\\
		Villasimius, Italy}

\begin{document}

\section{Introduction}
The observation of the $\Delta I=1/2$ rule and direct CP violation in kaon decays has attracted extensive theoretical study but both still lack a quantitative explanation. The running of the Wilson coefficients down to $\sim 2$  GeV which represents short distance physics can only explain a factor of 2, far less than the experimental factor of 25 $\Delta I=1/2$ rule enhancement. The remaining enhancement comes from hadronic matrix elements which requires non-perturbative treatment. On the other hand, direct CP violation in kaon decays serves a very important check of the standard model's CKM mechanism of CP violation. While experimentalists have measured Re$(\epsilon'/\epsilon)=1.65(26)\times10^{-3}$, with only 16\% error, there is no reliable theoretical calculation based on the Standard Model. Previous chiral perturbation theory based lattice QCD calculation using 2+1 dynamical domain wall fermion fails to give a conclusive result because of large systematic error~\cite{Li:2008kc}. Therefore, a direct lattice calculation of $K\rightarrow\pi\pi$ decay is extremely important to provide an explanation of the nature of $\Delta I=1/2$ rule and direct CP violation. This is a notoriously difficult calculation, but with the increasing advance of computing power, we want to show that it is now accessible.

In this paper, we try to do a direct, brute force calculation of the needed weak matrix elements. The isospin 0 $\pi-\pi$ final state involves disconnected graphs that make the calculation very difficult. For these graphs, the noise does not decrease with the increasing time separation of the source and sink, while the signal does. Therefore,  huge statistics is needed to get a clear signal. As a trial calculation, we do this on a relatively small lattice , so it is easier to collect large statistics. We concentrate on the study of the statistical uncertainty since it is the major difficulty of the problem. We will mainly report our results for the $\Delta I=1/2$ calculation in this paper. The inclusion of the $\Delta I=3/2$ part is for completeness; a much better calculation of the $\Delta I=3/2$ amplitude alone on a large lattice can be found in~\cite{Lightman2010}. In Section 2, we summarize our setup of the calculation. Then our $\pi-\pi$ scattering results are given in Section 3. Section 4 shows the details of the $K^0$ to $\pi\pi$ contractions, and the calculated results and conclusion are shown in Section 5.
%We get quite good signal for the disconnected graph of $\pi-\pi$ scattering and demonstrate clearly the attractive interaction of isospin 0 $\pi-\pi$ state and repulsion of isospin 2 state. All contractions related to $K^0\rightarrow\pi\pi$ decays are carefully listed and calculated. We obtain $Re(A_0)$ with $25\%$ error(statistic error only), and the $\Delta I=1/2$ enhancement factor of 6 in our kinematics point. 

\section{Computational Details}
Our calculation uses the Iwasaki gauge action($\beta=2.13$) and a 2+1 flavor($m_l=0.01$, $m_s=0.032$) domain wall fermion action, with space time volume $16^3\times32$, and $L_s=16$. The inverse lattice spacing for these lattices is determined to be 1.73(3)GeV, and the residual mass is $m_{res}=0.00308(4)$. The propagators are calculated on each of the 32 time slices using a Coulomb gauge fixed wall source (used for mesons), and a random wall source (used  to calculate loops in $type3$ and $type4$ graphs in Fig. \ref{Fig:contractions}). One propagator needs 12 (3 colors and 4 spins) Dirac operator inversions, so all together we carry out a few sets of 384 inversions for different sources and quark masses on a given configuration. This calculation is accelerated by a factor of 2-3 for $m_l=0.01$ by computing the Dirac eigenvectors with the smallest 35 eigenvalues and limiting the conjugate gradient inversion to the remaining orthogonal subspace. In order to obtain an on shell $K^0\rightarrow\pi\pi$ decay amplitude, the valence strange quark is partially quenched and its mass is chosen to be $m_s=0.066,0.099,0.165$, with the corresponding kaon mass shown in Tab.~\ref{Table:Mass}. In the following section, we will see that we can interpolate to on shell decay kinematics for both the $I=2$ and $I=0$ channels. This calculation is done on 400 configurations separated by 10 trajectories each.
%We use periodic boundary conditions, and consider the case that mesons have zero momentum or $P_x=2\pi/L$. 
%The energy of the particles(it is the mass when $\vec{P}=0$) are shown in Tab.~\ref{Table:Mass}.

\begin{table}
\caption{Masses of pion and kaons and energies of the two-pion states. Here $E_{I0}^\prime$ represents the isospin 0, two-pion energy when the disconnected graph V is ignored.}
\label{Table:Mass}

\begin{tabular*}{\textwidth}{@{\extracolsep{\fill}}lllllll}
\hline
\hline
$m_\pi$ & $E_{I0}$ & $E_{I0}^\prime$ & $E_{I2}$ & $m_k(0)$ & $m_k(1)$ & $m_k(2)$ \\
%P & $E_\pi$ & $E_{I0}$ & $E_{I0}^\prime$ & $E_{I2}$ & $E_k(0)$ & $E_k(1)$ & $E_k(2)$ \\
\hline
0.24267(68) & 0.450(17) & 0.4392(59)& 0.5054(15) & 0.4255(6) & 0.5070(6) & 0.6453(7) \\
%0 &	0.24267(68) & 0.450(17) & 0.4392(59)& 0.5054(15) & 0.4255(6) & 0.5070(6) & 0.6453(7) \\
%1 & 0.4698(35) & 0.753(25) & 0.6987(87) & 0.7382(39) & 0.5855(16) & 0.6485(15) & 0.7647(14) \\
\hline
\hline
\end{tabular*}

\end{table}

\section{Two-pion Scattering}
\begin{figure}[!htb]
\begin{tabular}{ll}
\includegraphics[width=0.5\textwidth]{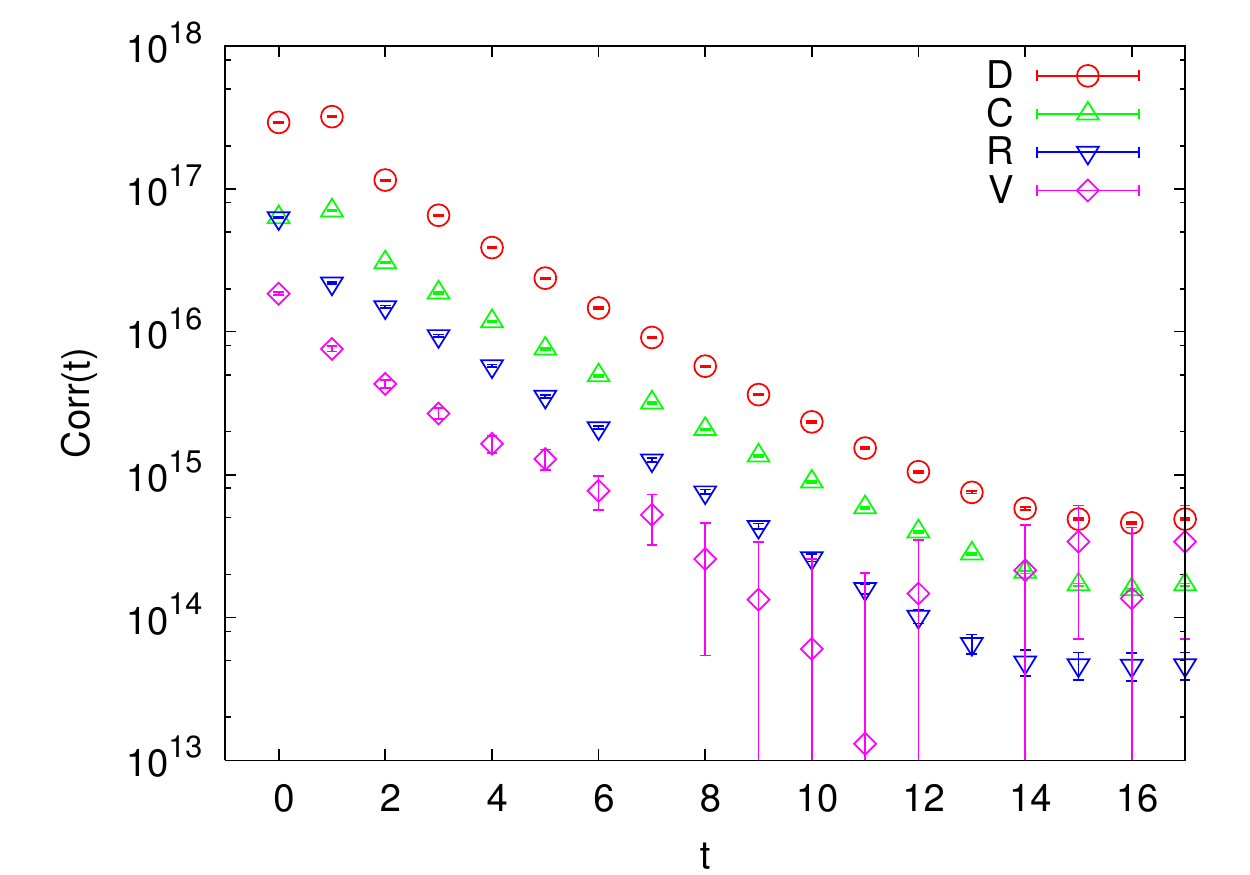} &
\includegraphics[width=0.5\textwidth]{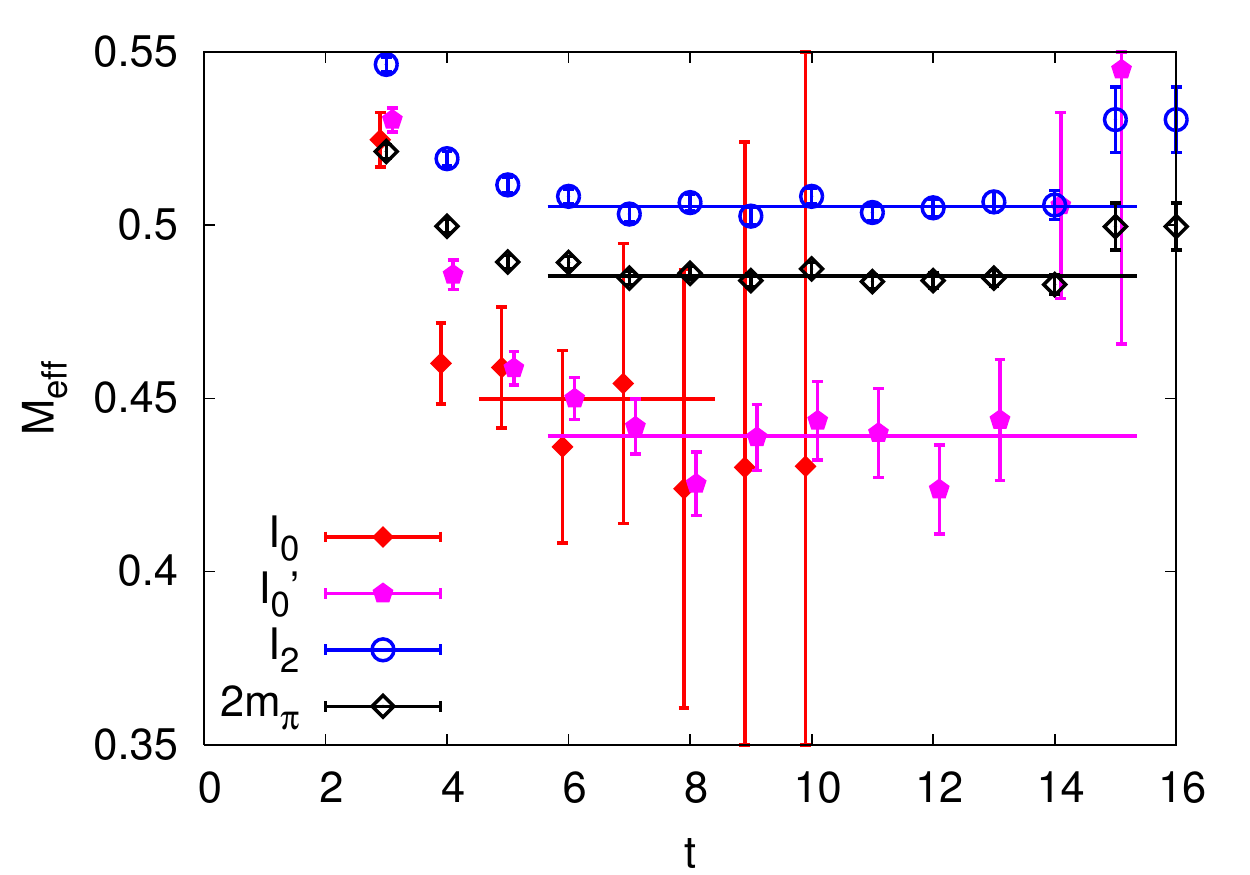} 
\end{tabular}
\caption{Left: the calculated results for the components of the correlation function defined as Direct(D), Cross(C), Rectangle(R), and Vacuum(V). Right: the effective mass for isospin 2 ($I_2$), isospin 0 ($I_0$), isospin 0 without the disconnected graph ($I_0^\prime$), and twice the pion effective mass ($2m_\pi$). }
\label{Fig:twopion}
\end{figure}

The $\pi-\pi$ scattering calculation includes 4 contractions, with the name Direct, Cross, Rectangle, and Vacuum diagram~\cite{Liu2009}. The calculated results from each of these four contractions are shown in the left panel of Fig.~\ref{Fig:twopion}. Notice that the disconnected (vacuum) graph has an almost constant error with increasing time separation% between the source and sink
 , so it appears to have an increasing error bar on the log plot,  while the signal exponentially decreases. The two-pion correlators are fit with a functional form Corr$(t)=|Z|^2 (exp(-E t)+exp(-E(T-t))+C)$, where the constant comes from the case in which the two pions propagate in different time directions. The fitted energies are summarized in Tab.~\ref{Table:Mass}. In order to see clearly the effect of the disconnected graph, we also do the calculation for the $I=0$ channel without the disconnected graph. These results are shown with labels that have an additional $\prime$ symbol. The right panel of Fig.~\ref{Fig:twopion} shows the resulting effective mass  for each case. It clearly shows the two pions are attractive in the $I=0$ channel and repulsive in the $I=2$ channel.

\section{$K^0$ to $\pi\pi$ Decay Contractions}

The effective hamiltonian for $K^0$ to $\pi\pi$ decay including the u, d, and s flavors as dynamic variables is
\begin{equation}
 H_w=\frac{G_F}{\sqrt{2}}V_{ud}^*V_{us}\sum_{i=1}^{10}[(z_i(\mu)+\tau y_i(\mu))] Q_i
\end{equation}
where the definition of the ten operators are the same as in~\cite{Blum2003}, $z_i$ and $y_i$ are the Wilson coefficients, and $\tau=-V_{ts}^*V_{td}/V_{ud}V_{us}^*$. To calculate the decay amplitudes $A_2$ and $A_0$, we need to calculate the weak matrix elements $\left<\pi\pi|Q_i|K^0\right>(a^{-1})$ on the lattice.

For simplicity, we list all possible contractions of $\left<\pi\pi|Q_i|K^0\right>$ in Fig.~\ref{Fig:contractions}. There are 48 different terms, label by circled numbers from 1 to 48, and grouped in terms of geometrical structure into $type1$, $type2$, $type3$, and $type4$. The calculation for the correlation functions of $\left<\pi\pi(t_\pi)|Q_i|(t)K^0(t_k)\right>$ is then straightforward, for example, in the I=0 (use notation $\left |I, I_z\right >$) case,  

\newcommand{\fig}[1]{\textcircled{{\scriptsize #1}}}
\begin{eqnarray}
%<00|Q_1|K^0> & = &i\frac{1}{\sqrt{3}}\{-\fig{1}-2\cdot\fig{5}+3\cdot\fig{9}+3\cdot\fig{17}-3\cdot\fig{33}\}\\
<00|Q_2|K^0> & = &i\frac{1}{\sqrt{3}}\{-\fig{2}-2\cdot\fig{6}+3\cdot\fig{10}+3\cdot\fig{18}-3\cdot\fig{34}\}\\
%<00|Q_3|K^0> & = &i\sqrt{3}\{-\fig{5}+2\cdot\fig{9}-\fig{13}+2\cdot\fig{17}+\fig{21}-\fig{25}\nonumber\\
%&&-\fig{29}-2\cdot\fig{33}-\fig{37}+\fig{41}+\fig{45}\}\\
%<00|Q_4|K^0> & = &i\sqrt{3}\{-\fig{6}+2\cdot\fig{10}-\fig{14}+2\cdot\fig{18}+\fig{22}-\fig{26}\nonumber\\
%&&-\fig{30}-2\cdot\fig{34}-\fig{38}+\fig{42}+\fig{46}\}\\
%<00|Q_5|K^0> & = &i\sqrt{3}\{-\fig{7}+2\cdot\fig{11}-\fig{15}+2\cdot\fig{19}+\fig{23}-\fig{27}\nonumber\\
%&&-\fig{31}-2\cdot\fig{35}-\fig{39}+\fig{43}+\fig{47}\}\\
<00|Q_6|K^0> & = &i\sqrt{3}\{-\fig{8}+2\cdot\fig{12}-\fig{16}+2\cdot\fig{20}+\fig{24}-\fig{28}-\fig{32}-2\cdot\fig{36}-\fig{40}+\fig{44}+\fig{48}\}
%<00|Q_7|K^0> & = &i\frac{\sqrt{3}}{2}\{-\fig{3}-\fig{7}+\fig{11}+\fig{15}+\fig{19}\nonumber\\
%&&-\fig{23}+\fig{27}+\fig{31}-\fig{35}+\fig{39}-\fig{43}-\fig{47}\}\\
%<00|Q_8|K^0> & = &i\frac{\sqrt{3}}{2}\{-\fig{4}-\fig{8}+\fig{12}+\fig{16}+\fig{20}\nonumber\\
%&&-\fig{24}+\fig{28}+\fig{32}-\fig{36}+\fig{40}-\fig{44}-\fig{48}\}\\
%<00|Q_9|K^0> & = &i\frac{\sqrt{3}}{2}\{-\fig{1}-\fig{5}+\fig{9}+\fig{13}+\fig{17}\nonumber\\
%&&-\fig{21}+\fig{25}+\fig{29}-\fig{33}+\fig{37}-\fig{41}-\fig{45}\}\\
%<00|Q_{10}|K^0> & = &i\frac{\sqrt{3}}{2}\{-\fig{2}-\fig{6}+\fig{10}+\fig{14}+\fig{18}\nonumber\\
%&&-\fig{22}+\fig{26}+\fig{30}-\fig{34}+\fig{38}-\fig{42}-\fig{46}\}
\end{eqnarray}
A few notes about the contractions shown in the Fig.~\ref{Fig:contractions}:
\begin{enumerate}
\item The graphs themselves do not carry the minus sign from the odd number of fermion loops. 
\item The dashed line stands for the contraction of colors. If there is no dashed line, it means that the trace of color is the same as the trace of spin.
\item A line stands for a light quark propagator if it is not explicitly labeled with 's'.
\item Using Fietz symmetry, it can be shown that there are 12 identities among these contractions, such as \fig{6}=-\fig{1}, \fig{5}=-\fig{2}.
\item Based on charge conjugation symmetry, the average of each of these contractions is real.
\item The loop contractions in $type3$ and $type4$ are calculated with Gaussian stochastic wall sources.
\end{enumerate}
Two examples of the definition of these graphs:
\begin{eqnarray}
\fig{1}&=&Tr\{\gamma_\mu(1-\gamma_5)L(x_{op},x_2)\gamma_5L(x_2,x_{op})\}\cdot Tr\{\gamma^\mu(1-\gamma_5)L(x_{op},x_1)\gamma_5L(x_1,x_0)\gamma_5S(x_0,x_{op})\}\nonumber\\
\fig{2}&=&Tr_c\{Tr_s\{\gamma_\mu(1-\gamma_5)L(x_{op},x_2)\gamma_5L(x_2,x_{op})\}\cdot Tr_s\{\gamma^\mu(1-\gamma_5)L(x_{op},x_1)\gamma_5L(x_1,x_0)\gamma_5S(x_0,x_{op})\}\}\nonumber
\end{eqnarray}
where $x_0$ is the position of the kaon, $x_1$ and $x_2$ are the position of the two pions, $L(x_{sink}, x_{src})$ is the light quark propagator, and $S(x_{sink},x_{src})$ is the strange quark propagator. \rm{Tr}$_c$ stands for color trace, \rm{Tr}$_s$ for spin trace, and \rm{Tr} for both spin and color trace.

Notice that the $type3$ and $type4$ graphs include quark loop integration which results in quadratic divergence. However, we should also notice that the operator renormalization allows the mixing with the lower dimensional operators $\bar{s}\gamma_5d$ and $\bar{s}d$~\cite{Blum2003}, where the latter one is forbidden by parity conservation. The subtraction of $\left<00|\bar{s}\gamma_5d|K^0\right>$ removes the quadratic divergence, and the subtracted results are calculated as
\begin{equation}
\left<00|Q_i|K^0\right>_{sub} = \left<00|Q_i|K^0\right> - \alpha_i \left<00|\bar{s}\gamma_5d|K^0\right>
\label{Eq:sub}
\end{equation}
where the subtraction coefficient $\alpha_i$ can be calculated from the $K^0$ to vacuum ratio $\frac{\left<0|Q_i|K^0\right>}{\left<0|\bar{s}\gamma_5d|K^0\right>}$.

\begin{figure}
  \begin{tabular*}{\textwidth}{@{\extracolsep{\fill}}cc}
%Color unmixed diagrams & Color mixed diagrams\\
\hline
%  &  \\
\includegraphics[width=0.4\textwidth]{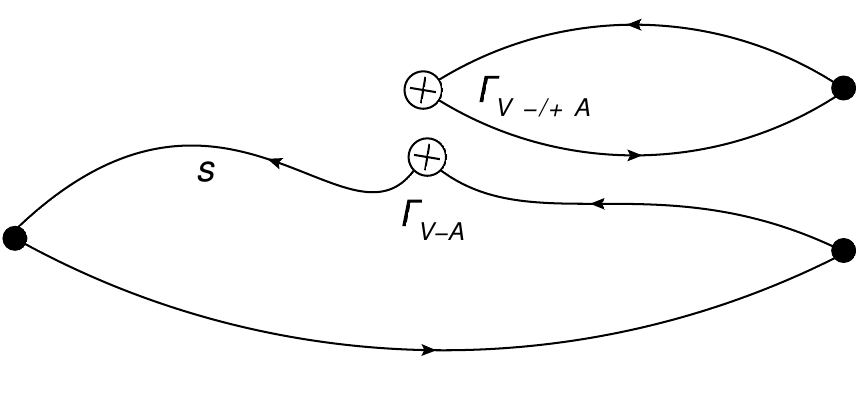}&
\includegraphics[width=0.4\textwidth]{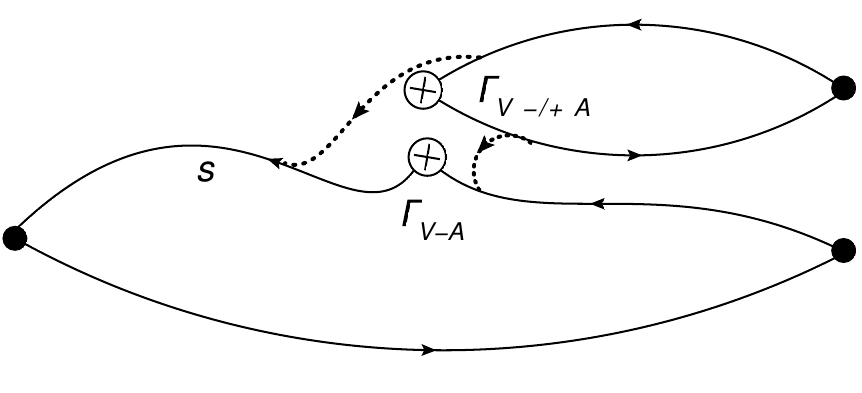}\\
%$1/3$ & $2/4$\\
\includegraphics[width=0.4\textwidth]{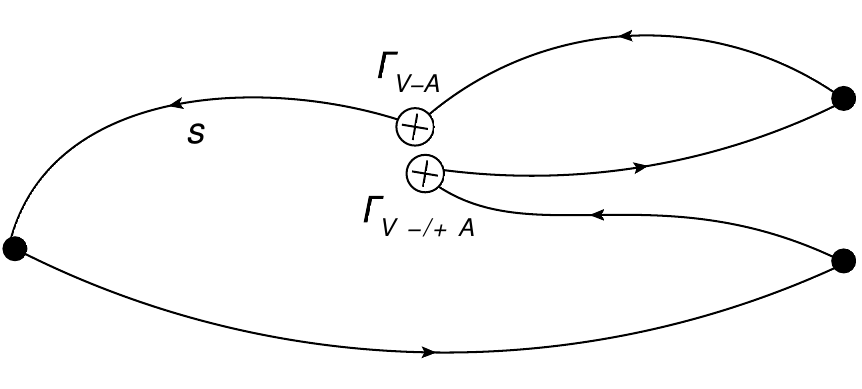}&
\includegraphics[width=0.4\textwidth]{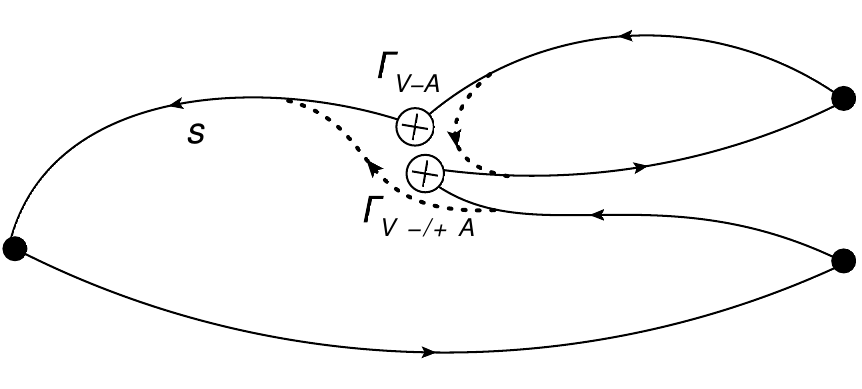}\\
%$5/7$ & $6/8$\\
\hline
%$\fig{6}=-\fig{1}$ & $\fig{5}=-\fig{2}$
%\end{tabular}
%\caption{type1}
%\label{Fig:type1}
%\end{figure}
%
%\begin{figure}[!htb]
%\begin{tabular}{c|c}
%Color unmixed diagrams & Color mixed diagrams\\
%\hline
%  &  \\
\includegraphics[width=0.4\textwidth]{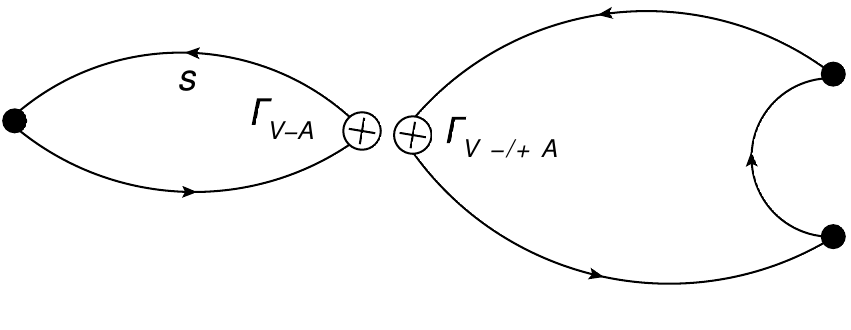}&
\includegraphics[width=0.4\textwidth]{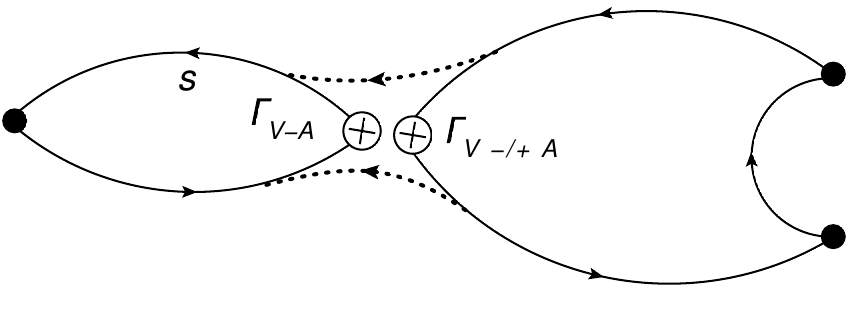}\\
%$9/11$ & $10/12$\\
\includegraphics[width=0.4\textwidth]{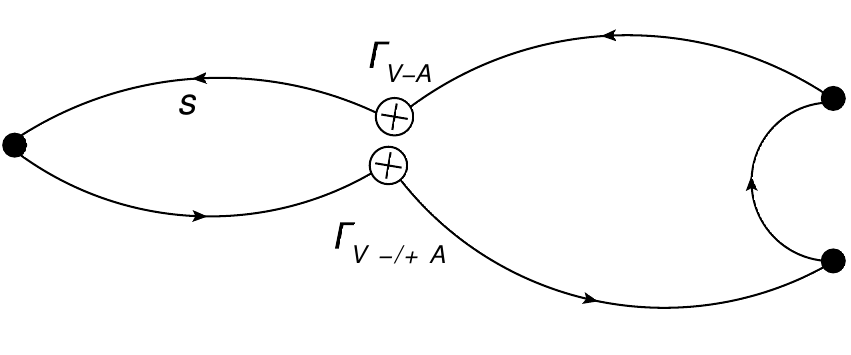}&
\includegraphics[width=0.4\textwidth]{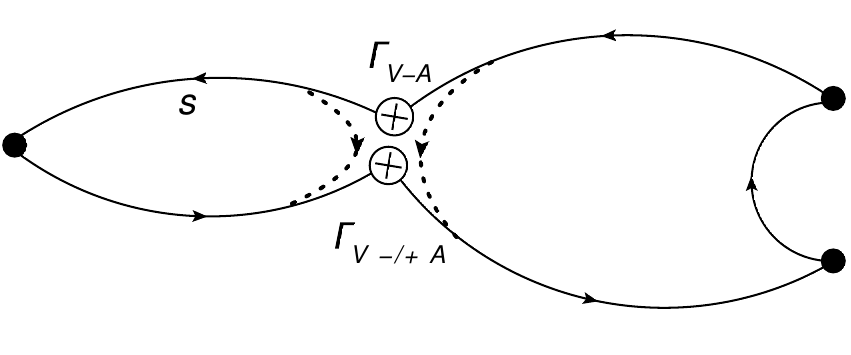}\\
%$13/15$ & $14/16$\\
\hline
%$\fig{14}=-\fig{9}$ & $\fig{13}=-\fig{10}$
%\end{tabular}
%\caption{type2}
%\label{Fig:type2}
%\end{figure}
%
%\begin{figure}[!htb]
%\begin{tabular}{c|c}
%Color unmixed diagrams & Color mixed diagrams\\
%\hline
%  &  \\
\includegraphics[width=0.4\textwidth]{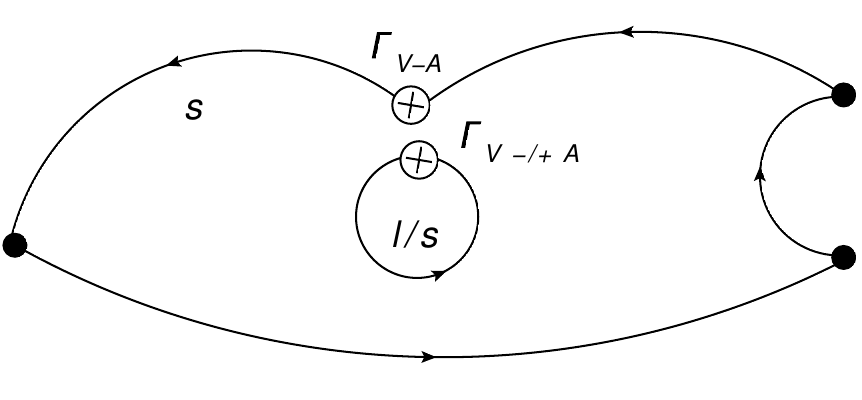}&
\includegraphics[width=0.4\textwidth]{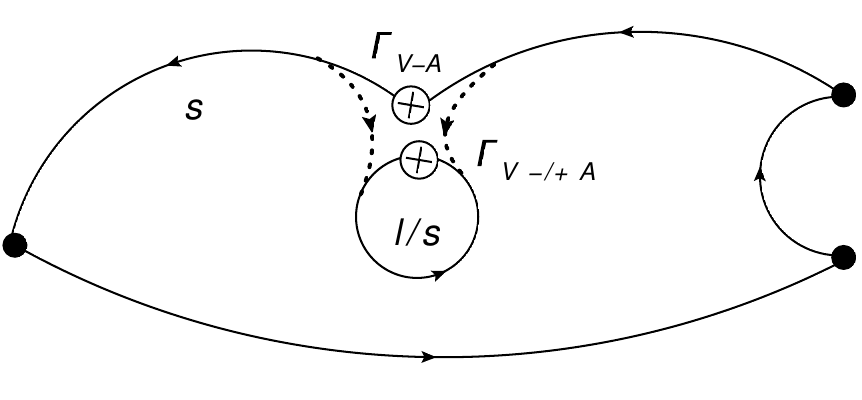}\\
%$17/19/21/23$ & $18/20/22/24$\\
\includegraphics[width=0.4\textwidth]{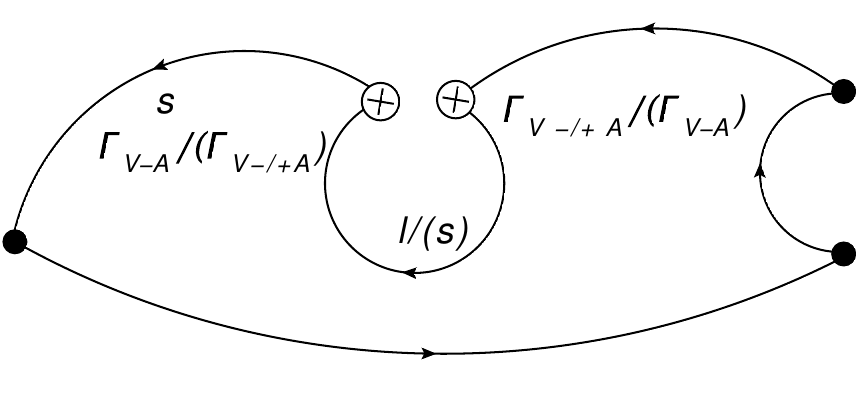}&
\includegraphics[width=0.4\textwidth]{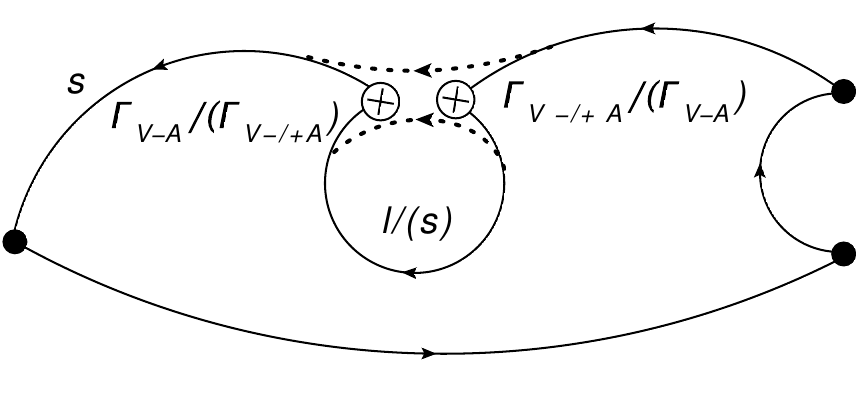}\\
%$25/27/29/31$ & $26/28/30/32$\\
\hline
%\end{tabular}
%\caption{type3}
%\label{Fig:type3}
%\end{figure}
%
%\begin{figure}[!htb]
%\begin{tabular}{c|c}
%Color unmixed diagrams & Color mixed diagrams\\
%\hline
%  &  \\
\includegraphics[width=0.4\textwidth]{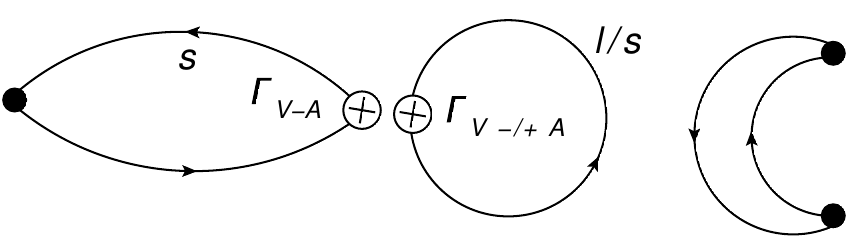}&
\includegraphics[width=0.4\textwidth]{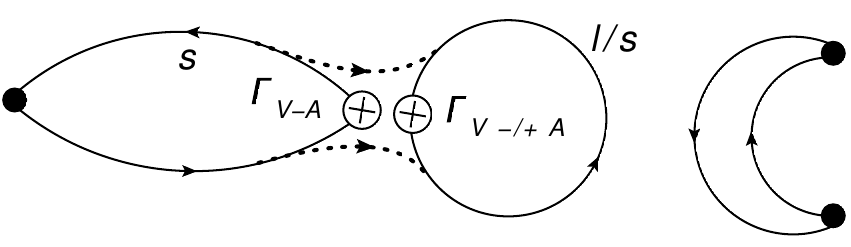}\\
%$33/35/37/39$ & $34/36/42/44$\\
\includegraphics[width=0.4\textwidth]{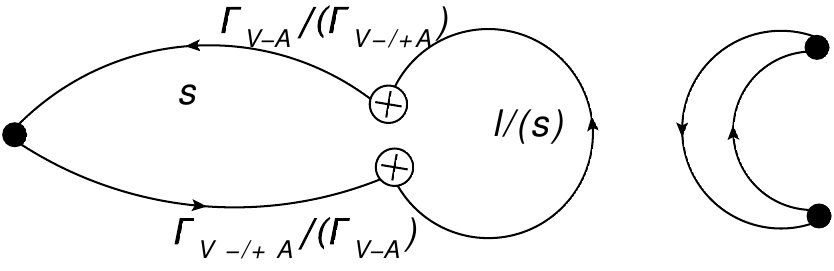}&
\includegraphics[width=0.4\textwidth]{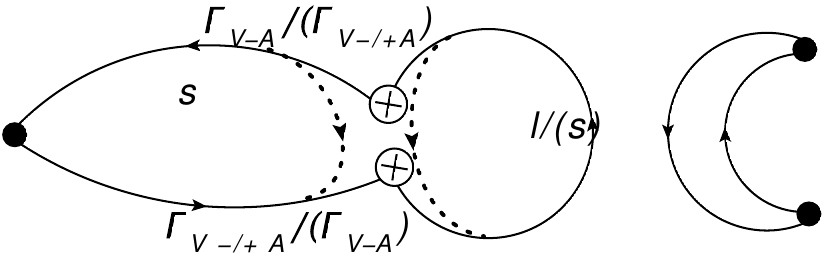}\\
%$41/43/45/47$ & $42/44/46/48$\\
\hline
\end{tabular*}
\caption{All $\left<\pi\pi|Q_i|K^0\right>$ contractions. They are labelled from left to right, top to bottom by the numbers 1 to 48. If there is a label '-/+', then it stands for two rows with the '-' sign comes first. If there is an additional label 'l/s', then it stands for 4 rows with the order '-l', '-s', '+l', '+s', e.g. the first graph is labelled by \fig{1} and \fig{3}.}
\label{Fig:contractions}
\end{figure}

The term $\left<00|\bar{s}\gamma_5d|K^0\right>$ comes from two contractions, one connected and one disconnected, which are labeled as $mix3$ and $mix4$ with the coefficent $\alpha_i$ incorporated. To better visualize the contributions from different types of contractions, we can write Eq.~\ref{Eq:sub} symbolically as
\begin{eqnarray}
\left<00|Q_i|K^0\right>_{sub} & = &type1 + type2 +type3 + type4 - mix3 - mix4 \nonumber \\
& = & type1 + type2 + sub3 + sub4
\end{eqnarray}
where $sub3=type3-mix3$, $sub4=type4-mix4$. 

\section{Results and Conclusions}
\begin{figure}[!htb]
\begin{tabular}{ll}
   \includegraphics[width=0.5\textwidth]{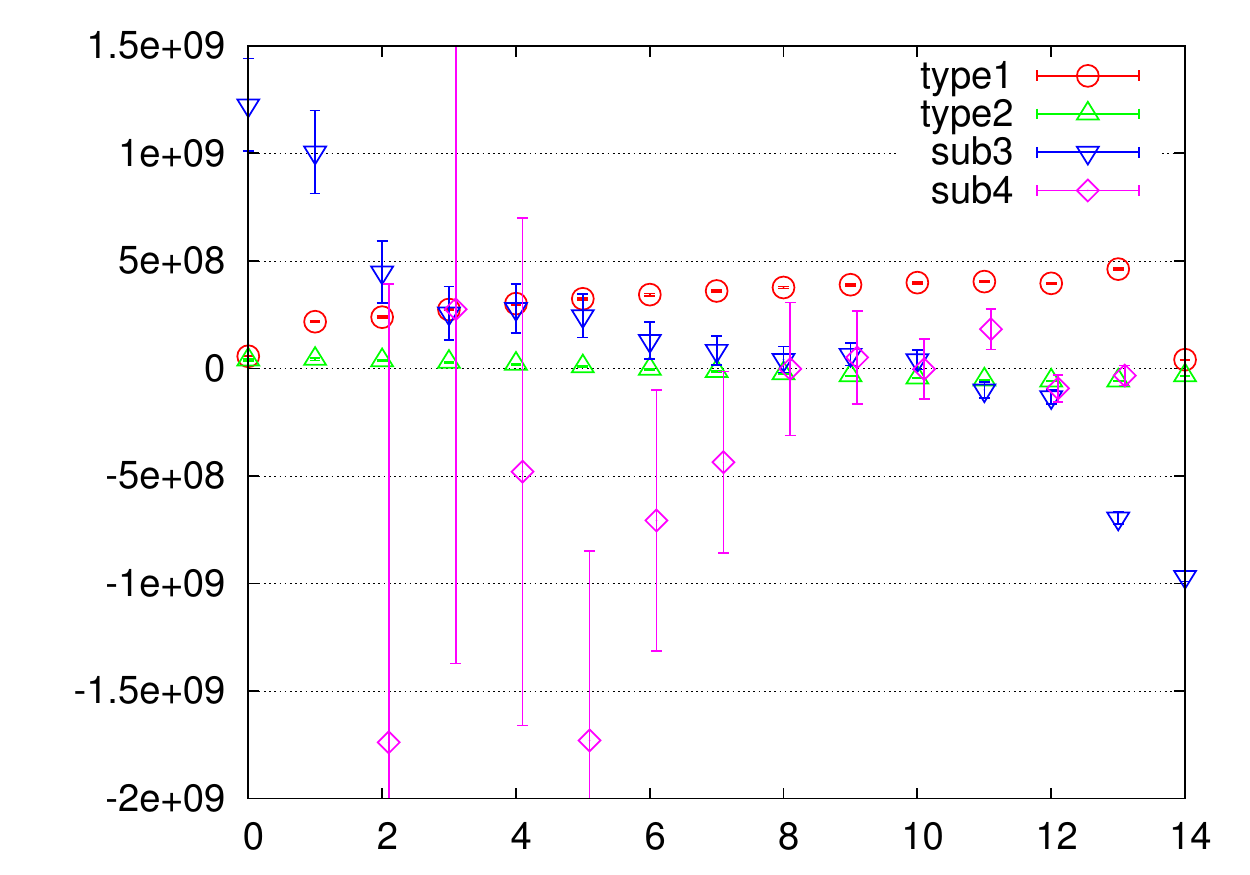} &
    \includegraphics[width=0.5\textwidth]{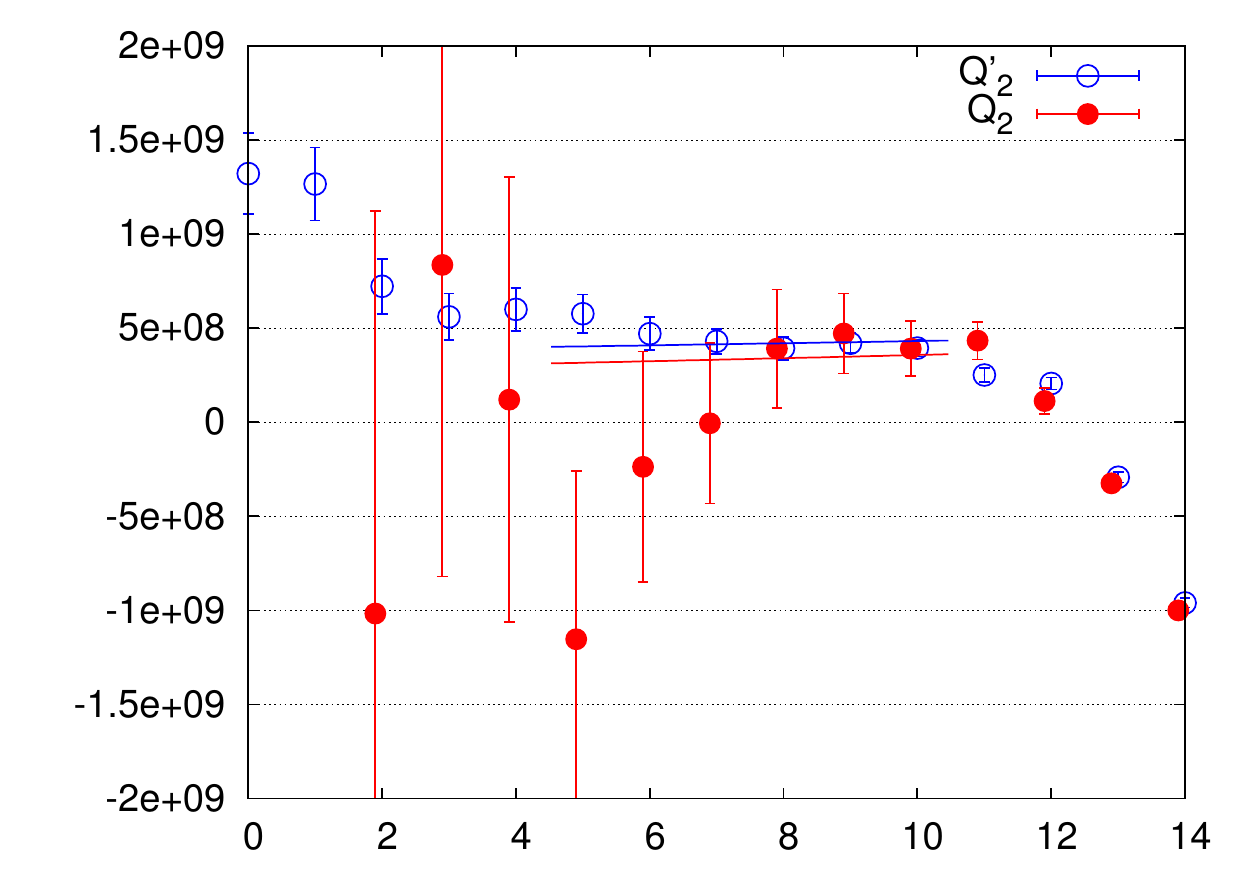} 
 \end{tabular}
 \caption{Left: Results of each type contraction of $\left<00(t_\pi=14)|Q_2(t)|K^0(t_k=0)\right>_{sub}$. Right: The total results of this correlation function and the fitting, where $Q_2$ labels the total result and $Q_2^\prime$ labels the result without the disconnected graph($sub4$).  }
\label{Fig:Q2}
\end{figure}

Figure~\ref{Fig:Q2} shows the calculated result for operator $Q_2$. The disconnected graph again makes huge contribution to the error.  We fit the $K^0$ to $\pi\pi$ correlators with a single free parameter $Q_i^{lat}(a)$:
\begin{equation}
       <\pi\pi(t_\pi)|Q_i(t)|K(0)>_{sub}=Q_i^{lat}(a) Z_{\pi\pi}^* Z_k e^{-E_{\pi\pi}t_{\pi}} e^{-(m_k-E_{\pi\pi})t}
\end{equation}
where $Z_k$ and $m_k$ are calculated from the correlator $\left<K(t)K(0)\right>$, and $Z_{\pi\pi}$ and $E_{\pi\pi}$ are calculated from the two-pion correlator. To see the effect of the disconnected graph, it is ignored and the calculated results are shown with an additional $\prime$ label. The fitted results are shown in Tab.~\ref{Tab:Qi}. 

\begin{table}
\centering
\caption{The fitted results for the weak matrix elements of $\Delta I=1/2$ kaon decay. The row label with a prime symbol means the disconnected graph is ignored. These are obtained using a source-sink separation of 14.}
\label{Tab:Qi}
\begin{tabular*}{\textwidth}{@{\extracolsep{\fill}}llllll}
  \hline
  \hline
i & 1 & 2 & 3 & 4 & 5 \\
\hline
$Q_i^{lat \prime} (\times 10^{-2})$ & -0.65(38) & 1.75(14) & 1.0(10) &  3.39(80) &  -5.04(91) \\
$Q_i^{lat}(\times 10^{-2})$ & -0.4(12) & 1.37(52) & 1.2(33) & 2.9(27) & -1.7(30) \\
\hline
\hline
i & 6 & 7 & 8 & 9& 10 \\
\hline
 $Q_i^{lat \prime} (\times 10^{-2})$ & -15.9(10) & 14.35(44) & 44.2(11) & -1.50(29) & 0.92(29) \\
$Q_i^{lat}(\times 10^{-2})$ & -6.4(40) & 11.6(12) & 34.9(24) & -1.0(10) & 0.66(97) \\
\hline
%\begin{tabular}{|l|l|lll|}
%\hline
%i & $Q_i^\prime(a)$ & $Q_i(a)$ &  $\%$ to $Re(A_0)$ & $\%$ to $Im(A_0)$\\
%\hline
%1  & -6.5(38)e-03  & -4(12)e-03 & \textcolor{blue}{8.5} & 0 \\
%2  & 1.75(14)e-02  & 1.37(52)e-02 & \textcolor{red}{91.6} & 0 \\
%3  & 1.0(10)e-02   & 1.2(33)e-02 & 0.003 & 6.8 \\
%4  & 3.39(80)e-02  & 2.9(27)e-02 & 0.50 & \textcolor{blue}{-61.1} \\
%5  & -5.04(91)e-02 & -1.7(30)e-02 & -0.03 & -2.7 \\
%6  & -1.59(10)e-01 & -6.4(40)e-02 & -0.37 & \textcolor{red}{141.2} \\
%7  & 1.435(44)e-01 & 1.16(12)e-01 & 0.02 & -0.48 \\
%8  & 4.42(11)e-01  & 3.49(24)e-01 & -0.14 & 9.6 \\
%9  & -1.50(29)e-02 & -1.0(10)e-02 & -0.0003 & 5.8\\
%10 & 9.2(29)e-03   & 6.6(97)e-03 & 0003 & 0.82 \\
\hline
\end{tabular*}
\end{table}

%\begin{figure}[!htb]
%\begin{tabular}{cc}
 %   \includegraphics[width=0.5\textwidth]{figures_400/Ave_Q6_mom_0_s_0_deltat_14_type3} &
%    \includegraphics[width=0.5\textwidth]{figures_400/Ave_Q6_mom_0_s_0_deltat_14_type4} \\
%    \includegraphics[width=0.5\textwidth]{figures_400/Ave_Q6_mom_0_s_0_deltat_14_All} &
 %   \includegraphics[width=0.5\textwidth]{figures_400/Ave_Q6_mom_0_s_0_deltat_14_fit} \\
%\end{tabular}
%\caption{k to pipi}
%\label{Fig:Q6}
%\end{figure}

The value of $Q_i(\mu)$ defined in the $\overline{MS}$ scheme can be calculated from $Q_i(\mu)=Z_{ij} Q_j^{lat}(a)$, where the Non-Perturbative Renormalization(NPR) matrix $Z_{ij}$ and the choice of the basis follow closely those in \cite{Blum2003, Li2008}. 
%The values of the NPR factor $Z_{ij}(2.15GeV)$ and the Wilson coefficients $y_i$(2.15GeV), $z_i$(2.15GeV) are taken from \cite{Li2008}. 
The finite volume effect are removed by the Lellouch-L\"uscher factor (F) \cite{Lellouch2001} which relates the quantity $M$ calculated from the lattice in finite volume to the infinite volume result $A$ based on our convention of phase space factor:
\begin{equation}
|A|^2 = 4\pi(\frac{E_{\pi\pi}^2m_K}{p^3})\{p\frac{\partial\delta(p)}{\partial p} + q\frac{\partial\phi(q)}{\partial q}\}|M|^2 = F^2 |M|^2
\label{Eq:LL}
\end{equation}
where $p$ is defined in $E_{\pi\pi}=\sqrt{m_\pi^2+p^2}$, and $q=Lp/2\pi$.
Taking the free field limit, this becomes $|A|^2 = 2(2m_\pi)^2 m_K L^3 |M|^2$, and the pre-factors show the different normalization of states in a box and states in infinite volume. For the isospin 0 state in our case, $p^2$ has a large negative value. Even though Eq.~\ref{Eq:LL} can be analytically continued to imaginary momentum $p$, it does not work well for large negative $p^2$ as the function $\phi(q)$ becomes ill defined. We believe that this difficulty results because the condition on the interaction range $R<L/2$ used to derive the Lellouch-L\"uscher factor is not well satisfied. Therefore, it is safer for us to use the free field factor in the $I=0$ case and concentrate on the statistical error from the calculation of the correlators. This problem will naturally go away once we explicitly give momentum to the two pions or work in a larger volume.

Combining everything together, we are ready to calculate the $K^0$ to $\pi\pi$ decay amplitudes,
\begin{equation}
A_I e^{i\delta_I} = F \frac{G_F}{\sqrt{2}}V_{ud}V_{us}\sum_{i=1}^{10}[(z_i(\mu)+\tau y_i(\mu)) Z_{ij}(\mu) Q_i^{lat}(a^{-1})] 
\end{equation}
The calculated Re($A_0$) and Im($A_0$) are shown in Tab.~\ref{Tab:A}. For comparison, the calculated $\Delta I=3/2$ on shell  decay amplitudes of $m_k=0.5070(6)$ to $E_{I2}=0.5054(15)$ are
 Re($A_2$)=$5.394(45)\times 10^{-8}$ GeV and   Im($A_2$)$=-0.7792(78)\times 10^{-12}$ GeV.

\begin{table}[!tbh]
\caption{$K^0$ to $\pi\pi$ $\Delta I=1/2$ Decay amplitudes in unit of GeV.}
\label{Tab:A}
\begin{tabular*}{\textwidth}{@{\extracolsep{\fill}}lllllll}
\hline
$m_K$  & $Re(A^\prime_0)(\times 10^{-8})$ & $Re(A_0)(\times 10^{-8})$ & $Im(A^\prime_0)(\times 10^{-12})$ & $Im(A_0)(\times 10 ^{-12})$ \\
%& $Re(A_2)(\times 10^{-8}GeV)$ &  $Im(A_2)(\times 10 ^{-12}GeV)$ 
\hline 
0.4255(6) & $37.8(2.1)$ & $28.3(7.8)$ & $-62.1(5.2)$ & $-21(20)$ \\
0.5070(6) & $43.5(2.4)$ & $35.4(9.9)$ & $-67.7(5.5)$ & $-48(27)$ \\
\hline
on shell &  $38.7(2.1)$ & $30.4(8.5)$ & $-63.1(5.3)$ & $-29(22)$\\
\hline
\end{tabular*}
\end{table}

In conclusion, our zero momentum $K^0$(778 MeV) to $\pi$(420 MeV)$\pi$(420 MeV) decay calculation gives Re($A_0$) with an error of $25\%$. We find a ratio of Re($A_0$)/Re($A_2$) of roughly 6. Since our pion mass is so much heavier than the physical pion mass, a much smaller factor than the experimental value of 25 is expected. To have a definite conclusion on Im($A_0$) from which we could calculate $\epsilon^\prime$, we estimate that 4 times statistics could establish a non-zero value if we believe that the  true result is not too far away from the result without the disconnected graphs.

{\bf Acknowledgements}
I thank all my colleagues in the RBC and UKQCD collaborations for discussions, suggestions, and help. I especially thank Norman Christ for detailed instructions and discussions, and Ran Zhou for the deflation code. I acknowledge Columbia University, RIKEN, BNL and the U.S. DOE for providing the facilities on which this work was performed. This work was supported in part by U.S. DOE grant DE-FG02-92ER40699.

\bibliography{citations}
\bibliographystyle{h-physrev}

\end{document}